\documentclass{ws-ijmpd}
\usepackage{subfigure}
\begin{document}

\markboth{Liang et al.}
{Optical Emission Components of GRBs}

\catchline{}{}{}{}{}
%
%%%%%%%%%%%%%%%%%%%%%%%%%%%%%%%%%%%%%%%%%%%%%%%%%%%%%%%%%%%%%%%%%%%%

\title{Statistical Properties of Multiple Optical Emission Components in Gamma-Ray Bursts and Implications\footnote{This work is supported by the National
Natural Science Foundation of China under grants No. 11025313, 10873002, 11063001, 11163001 and 10847003, the National Basic Research Program (``973"
Program) of China (Grant 2009CB824800), Special Foundation for Distinguished Expert Program of Guangxi
, the Guangxi SHI-BAI-QIAN project (Grant 2007201), the Guangxi Natural Science Foundation
(2010GXNSFA013112 and 2010GXNSFC013011), the special funding for national outstanding young scientist (Contract No. 2011-135), and the 3th Innovation Projet of Guangxi University. BZ acknowledges support from NASA (NNX10AD48G) and NSF (AST-0908362).}}

\author{En-Wei Liang$^{1,2, 3}$, Liang Li$^{1}$, Qing-Wen Tang$^{1}$, Jie-Min Chen$^{1}$ and Bing Zhang$^{3}$}

\address{$^{1}$Department of Physics and GXU-NAOC Center for Astrophysics and Space
Sciences, Guangxi University, Nanning 530004, China. E-mail:lew@gxu.edu.cn\\
$^2$National Astronomical Observatories, Chinese Academy of Sciences, Beijing, 100012, China\\
$^{3
}$Department of Physics and Astronomy, University of Nevada, Las Vegas, NV 89154}
\maketitle

\begin{history}
\received{15 October 2011}
\revised{15 November 2011}
\comby{Managing Editor}
\end{history}

\begin{abstract}
Well-sampled optical lightcurves of 146 gamma-ray bursts (GRBs) are complied from the literature. Multiple optical emission components are extracted with  power-law function fits to these lightcurves. We present a systematical analysis for statistical properties and their relations to prompt gamma-ray emission and X-ray afterglow for each component. We show that peak luminosity in the prompt and late flares are correlated and the evolution of the peak luminosity may signal the evolution of the accretion rate. No tight correlation between the shallow decay phase /plateau and prompt gamma-ray emission is found. Assuming that they are due to a long-lasting wind injected by a compact object, we show that the injected behavior favors the scenarios of a long-lasting wind powered by a Poynting flux from a black hole via the Blandford-Znajek mechanism fed by fall-back mass or by the spin-down energy release of a magnetar after the main burst episode. The peak luminosity of the afterglow onset is tightly correlated with $E_{\gamma,\rm iso}$, and it is dimmer as peaking later. Assuming that the onset bump is due to the fireball deceleration by the external medium, we examine the $\Gamma_0-E_{\gamma, iso}$ relation and find that it is confirmed with the current sample. Optical re-brightening is observed in 30 GRBs in our sample. It shares the same relation between the width and the peak time as found in the onset bump, but no clear correlation between $L_{R, p}$ and $E_{\gamma, iso}$ similar to that observed for the onset bumps is found. Although its peak luminosity also decays with time, the slope is much shallower than that of the onset peak, as is the case for the onset bumps. We get $L\propto t^{-1}_{\rm p}$, being consistent with off-axis observations to an expanding external fireball in a wind-like circum medium. Therefore, the late re-brightening may signal another jet component. Mixing of different emission components may be the reason for the observed chromatic breaks of the shallow decay segment in different energy bands.
\end{abstract}

\section{Introduction}
Gamma-ray bursts (GRBs) and their broadband afterglows are the most luminous phenomenon in the Universe. The most popular model is the internal + external shock fireball model, which suggests that the prompt emission is produced by internal shocks at a distance internal to the deceleration radius of the GRB fireball and the broadband afterglows are from the external shocks when the fireball is decelerated by the ambient medium\cite{Rees1994,Meszaros1997,Sari1998}. Prompt and afterglow emissions involve two distinct processes at different sites in this model. The observations with {\em Swift} mission significantly improved our understanding of the internal+external shock picture for GRBs\cite{Zhang2007a}. Rapid localization with {\em Swift} also revolutionized the ground-based follow-up observations, leading to establishing a large sample of GRBs with well-sampled optical afterglow lightcurves and redshift measurement. Following our comprehensive analysis of the {\em Swift} data\cite{Liang2006a,Zhang2007b,Liang2007,Liang2008,Liang2009}, we make a systematical analysis of the optical data and explore their relations to the prompt gamma-ray emission and the X-ray afterglow.
\section{Data\label{sec:data}}
We include all the GRBs that have optical afterglow detection in our sample. A sample of 225 optical lightcurves are complied from the literature in the period from Feb. 28, 1997 to November 2011. We make an extensive search for the optical data from published papers or from GCN Circulars in case of no published paper being available for some GRBs. Well-sampled lightcurves are available for 146 GRBs.  We collect the optical spectral index ($\beta_O)$\footnote{An optical spectral index $\beta_O=0.75$ is used for those GRBs without $\beta_O$ available} and the extinction $A_{\rm V}$ by the host galaxy of each burst from the same literature. Galactic extinction correction is made by using a reddening map presented by Schlegel et al. (1998)\cite{Schlegel1998}. Since the $A_{\rm V}$ values are available only for some GRBs and the $A_{\rm V}$ is derived from the spectral fits by using different extinction curves, we do not make correction for the extinction of the GRB host galaxy. The $k$-correction in magnitude is calculated by $k=-2.5(\beta_O-1)\log(1+z)$. Note that most of the well-sampled optical lightcurves are in the R band. For a few GRBs, the data are well-sampled in the other bands. We correct these lightcurves to the R band with the optical spectral indices. For data in late epochs ($\sim 10^6$ seconds after the GRB trigger), possible flux contribution from the host galaxy is also subtracted. The isotropic gamma-ray energy ($E_{\rm \gamma, iso}$) is derived in the energy band of $1-10^4$ keV in the burst local frame with spectral indices. Data and references of our full sample will be reported in a series of papers (in preparation).
\section{Lightcurve Fitting and a Synthesized Optical Emission Lightcurve \label{sec:data}}
The optical lightcurves are usually composed of one or several power-law segments as well as flares/re-brightening features. The mix of different components makes the diversity of the optical afterglow lightcurves. Different from previous statistical analysis on the optical data by some teams\cite{Liang2006b,Panaitescu2008,Panaitescu2011,Kann2010,Kann2011}
we fit the lightcurves with a model of several components in order to subtract each emission component from the lightcurves. The basic components in our model are a power-law function and a smooth broken power-law, i.e.,
\begin{eqnarray}
&F=F_0 t^{-\alpha},\   \   \   \
&F=F_0 [(t/t_{\rm b})^{\alpha_1\omega}+(t/t_{\rm b})^{\alpha_2\omega}]^{-1/\omega},
\end{eqnarray}
where $\alpha$ is the temporal decay slope, $t_{\rm b}$ is the break (or the peak) time, and $\omega$ measures the sharpness of a break (or a peak) of a lightcurve.
The width of a flare/bump is measured with the full-width-at-half-maximum (FWHM). We develop an IDL code to make the best fit with a subroutine called MPFIT\footnote{http://www.physics.wisc.edu/~craigm/idl/fitting.html}. Note that the parameter $\omega$ is usually fixed at 3 or 1 in our fitting. The approach of our lightcurve fitting is as follows. Initially, we add components to our model by inspecting the global feature of a lightcurves. If the reduced $\chi^2_{\rm r}$ is much larger than 1, we continue to add components and make the fit. We try to get a fit with $\chi^2_{\rm r}$ being close to 1. The $\chi^2_{\rm r}$ values for some lightcurves are much lower than 1, indicating that some model parameters are poorly constrained. Therefore, we fix some parameters to make the fits for these GRBs. The erratic fluctuation of some data points with a small error bar in some GRBs, such as GRB 030329, makes $\chi^2_{\rm r}$ be much larger than 1. We do not add additional component for these data points and the $\chi^2_{\rm r}$ of the fits for these GRBs are much larger than 1. The most difficult problem of our fit is extraction of the seriously overlapped flares/bumps from the lightcurves. The slopes of these flares/bumps are usually quite uncertain. In our fitting, we first let all parameters free to get the best fit for the global lightcurve, then adjust the rising slopes to ensure that the fitting curve crosses the data point around the peak times of the two components. Finally, we fix the rising slopes and make the best fit again.

\begin{figure}
$
\begin{array}{lr}
\subfigure [A synthesized cartoon lightcurve of multiple optical emission components based on our statistics. (b) Examples of the lightcurves with various emission components.  The solid lines represent the best fit to the data. Simultaneous X-ray data
observed with {\em Swift}/XRT (crosses with error bars) are also presented.]{\includegraphics[angle=0,scale=0.55,height=1.6 in]{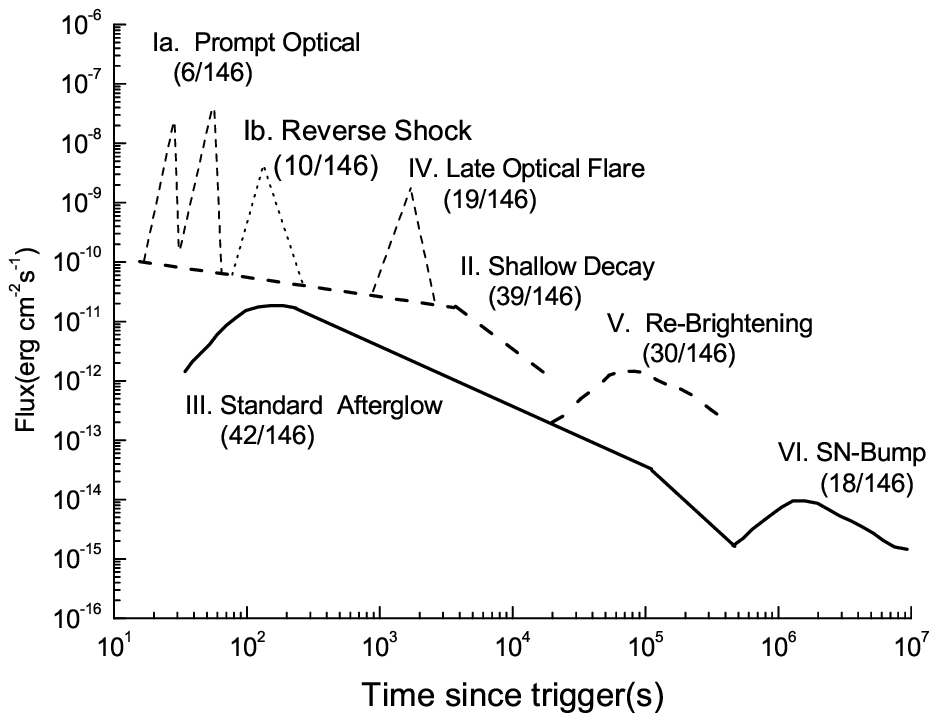}}&
\subfigure []{
 $
    \begin{array}{cc}
            \includegraphics[angle=0,scale=0.25,height=1.3 in]{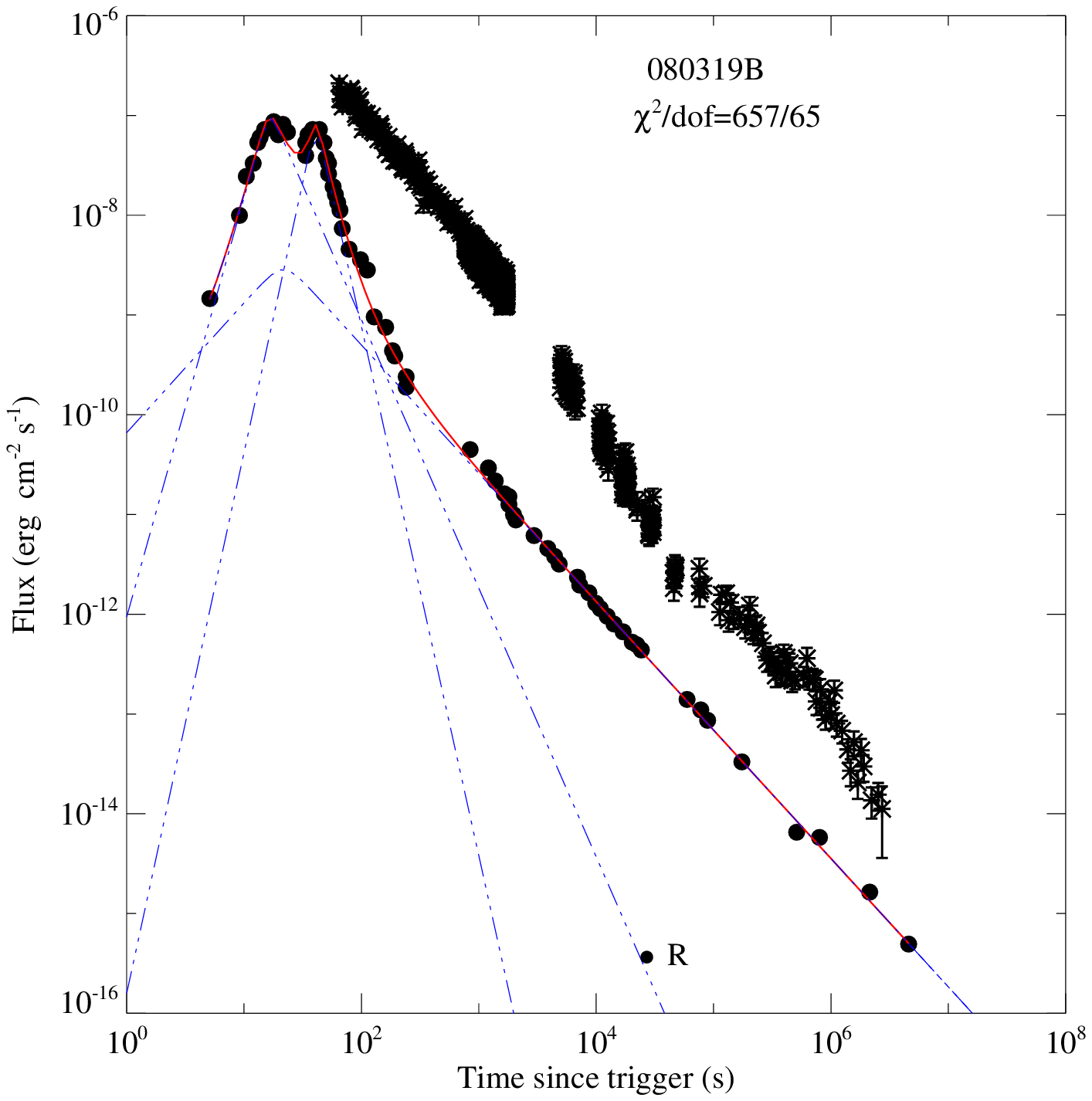}&
            \includegraphics[angle=0,scale=0.25,height=1.3 in]{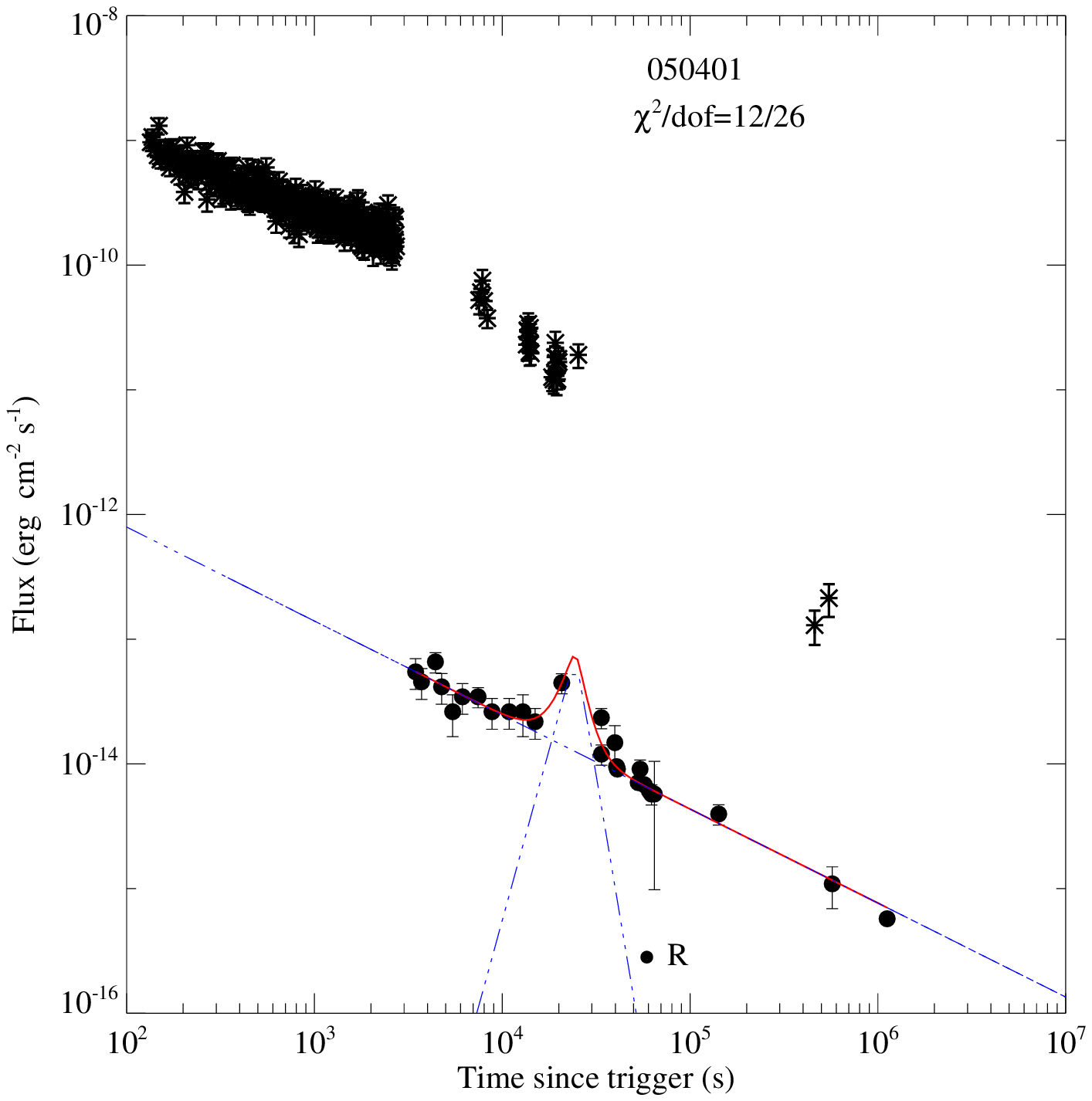}\\
            \includegraphics[angle=0,scale=0.25,height=1.3 in]{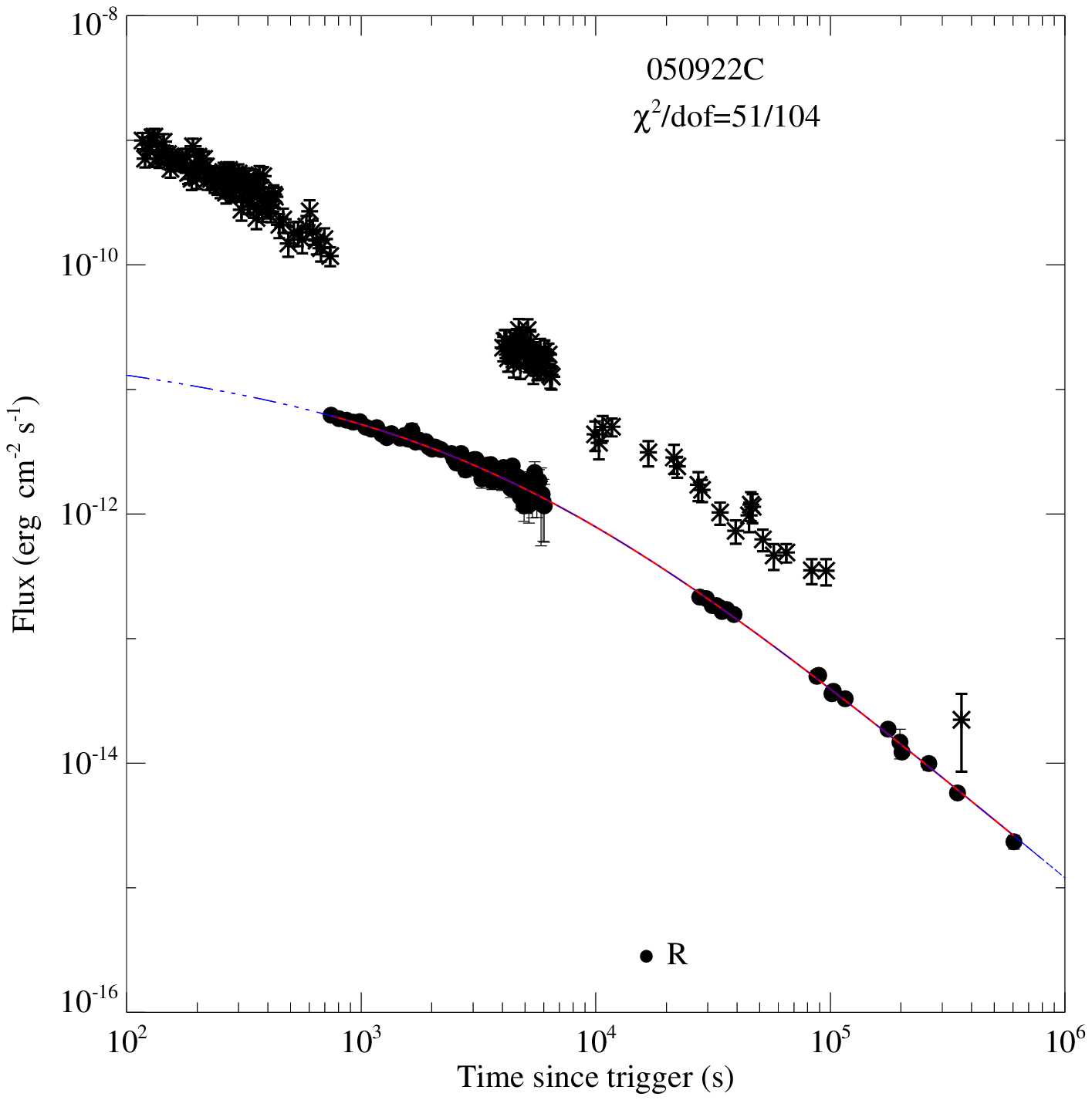}&
            \includegraphics[angle=0,scale=0.25,height=1.3 in]{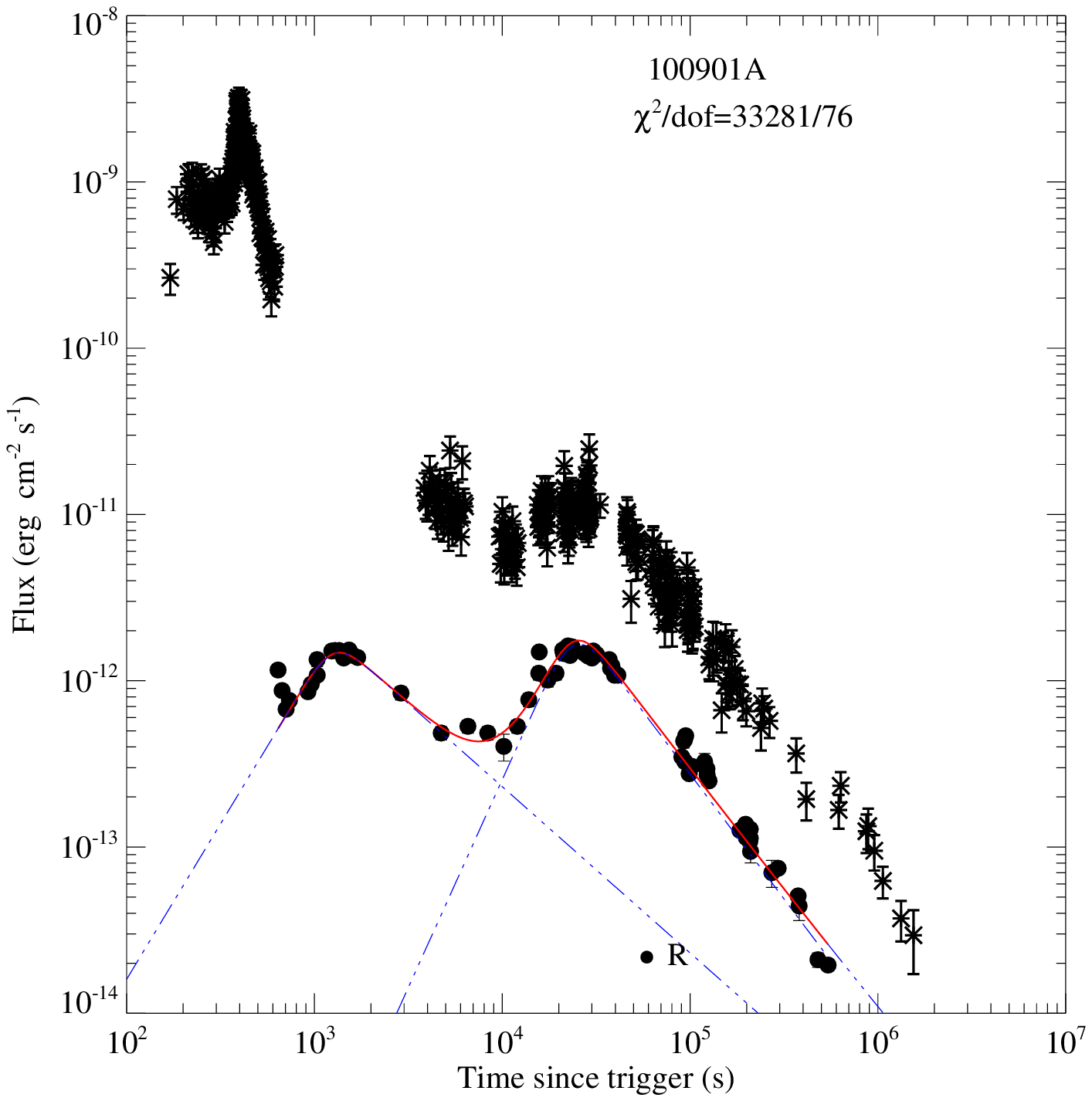}\\
    \end{array}
$
}
\end{array}
$
\caption{} \label{Cartoon}
\end{figure}

The flux of an flare event usually rapidly increases and drops. We identify a flare event with a criterion that the slopes of the rising and decaying parts are steeper than 2. Those optical flares during the burst duration are defined as prompt optical flares. Reversed shock flares following the prompt flares are observed in only a few GRBs. They are extensively discussed in literature. The standard fireball model suggests that the decay slope of the afterglows should be steeper than 0.75 if no any late energy injection after the GRB phase. We therefore define the shallow decay phase with the criterion that the initial decay slope of this segment is shallower than 0.75, which transits to a steeper decay after the break time. An optical afterglow onset is defined as an initial smooth hump with a peak less than 1 hour after the GRB trigger and decaying as a power-law with a slope being consistent with the external shock models ($0.75<\alpha<2$). A re-brightening hump is analogous to the afterglow onset hump but is later than the onset hump. The supernovae bumps are identified as those late re-brightenings that peak at around $1\sim 2$ weeks post the GRB trigger. We decompose emission components and make statistical analyses for each one. Morphologically, a synthesized optical lightcurve of these components along with some typical examples is shown in a cartoon picture (Figure \ref{Cartoon}) based our statistical results. It describes 7 components related to the X-ray canonical lightcurve\cite{Zhang2006}. The early optical afterglow lightcurves ($t<10^3$ second post the GRB trigger) of about one-third GRBs show a smooth bump and the other one-third lightcurves start with a shallow decay segment. Twenty-four optical flares are observed in 19 GRBs. Late re-brightenning is observed in 30 GRBs. A jet like break, in which the decay slope transits from $0.75\sim 1.5$ to $1.5\sim 2.5$, is detected in 10 GRBs\footnote{We do not include those breaks that have a slope shallower than 0.75 breaking to a slope steepen than 1.5}. A clear Supernova bump is detected for 18 GRBs. The detection probability of each component is also marked in the cartoon lightcurve. We report our results for the flares, shallow decays, afterglow onset bumps, and late re-brightening in this paper. We mark the parameters of these components with superscripts ``F" for the flares, ``S" for the shallow decays, ``A" for the afterglow, and ``R" for the re-brightening.

\section{Flares}
\begin{figure*}
\includegraphics[angle=0,scale=0.4,width=0.32\textwidth,height=0.25\textheight]{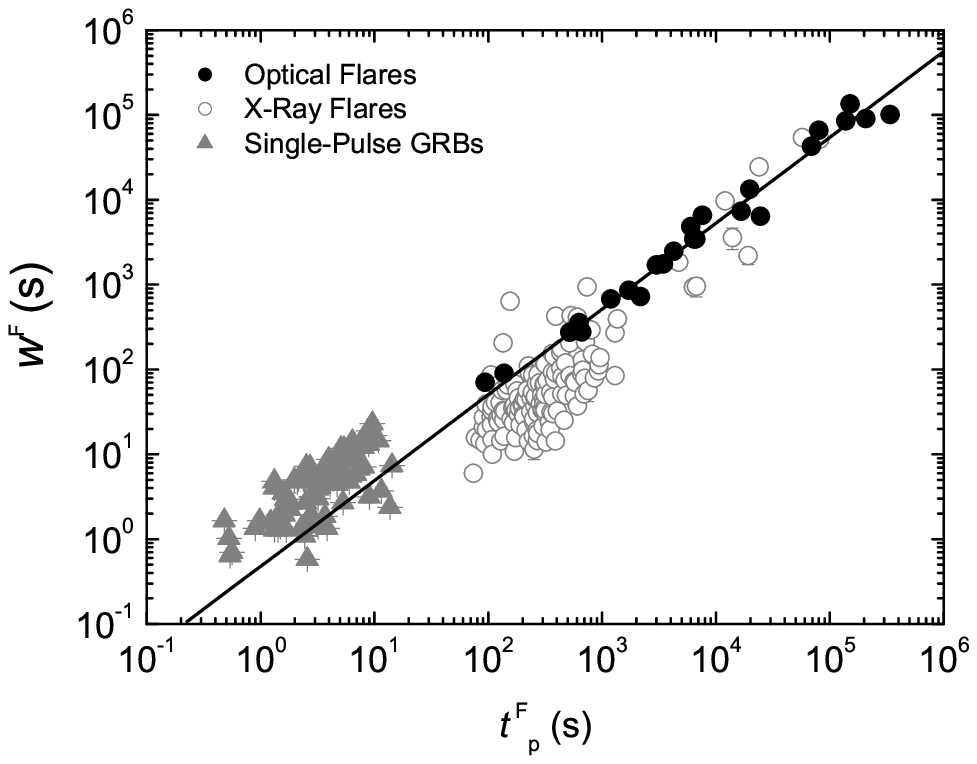}
\includegraphics[angle=0,scale=0.4,width=0.32\textwidth,height=0.25\textheight]{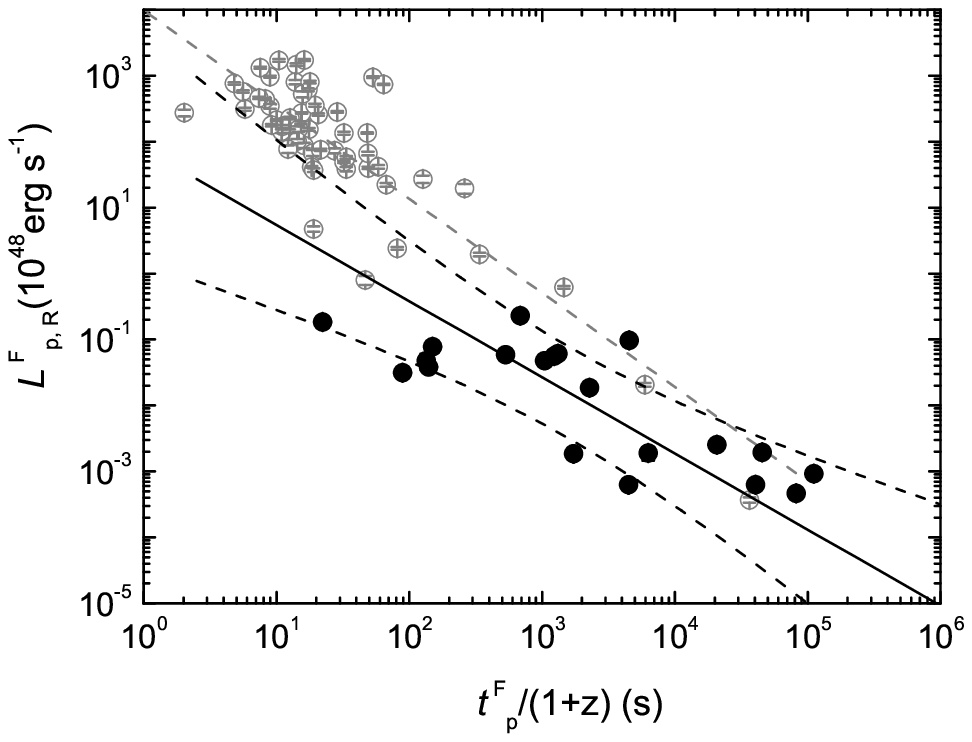}
\includegraphics[angle=0,scale=0.4,width=0.32\textwidth,height=0.25\textheight]{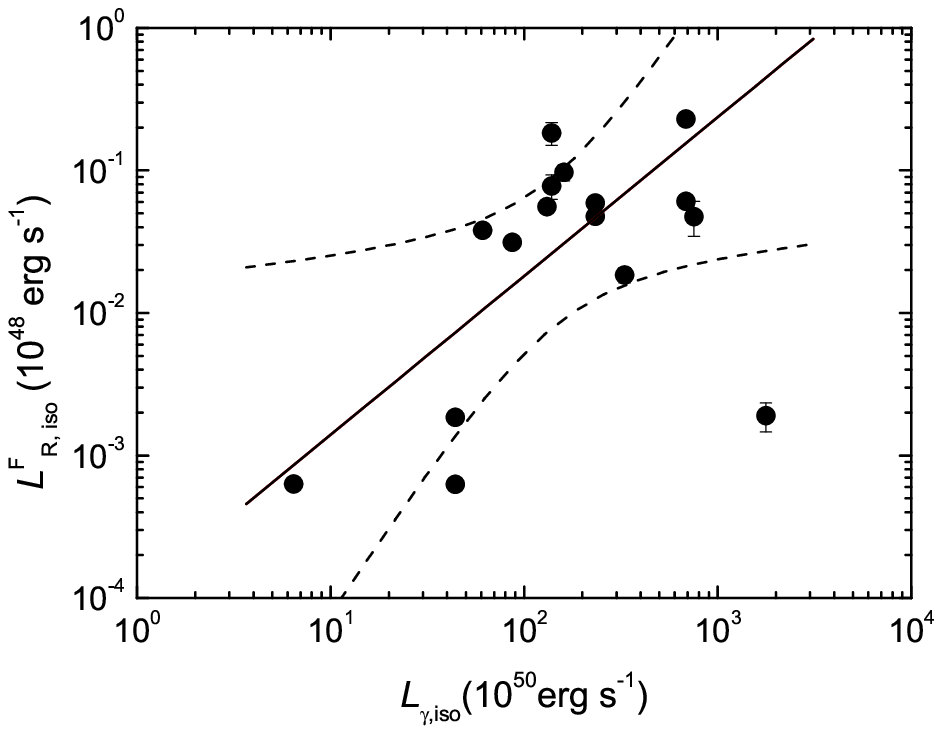}
\caption{Correlation between $w^{\rm F}$ and $t^{\rm F}_{\rm p}$ as well as relations of $L^{\rm F}_{\rm R, iso}$ to $t^{\rm F}_{\rm p}$ and $L_{\gamma, \rm iso}$. The black solid dots, grey open circles, and grey triangles are for the optical flares, X-ray flares, and prompt gamma-ray pulses. Best fit lines with $3\sigma$ significance level are also shown. } \label{Flare_Corr}
\end{figure*}

We get 24 late flares in 19 GRBs. Relations of the width ($w^{\rm F}$) and the peak luminosity ($L^{\rm F}_{\rm p,iso}$) of the flares as a function of the peak time ($t^{\rm F}_{\rm p}$) are shown in Figure \ref{Flare_Corr}. The $t^{\rm F}_{\rm p}$ ranges from $\sim$ tens of seconds to $\sim 10^6$ seconds. The $w^{\rm F}$ values are in the same range as $t^{\rm F}_{\rm p}$. The $L^{\rm F}_{\rm R, iso}$ ranges in $10^{43}-10^{49}$ erg s$^{-1}$, with a typical value of $10^{46}$ erg s$^{-1}$. A tight correlation between $w^{\rm F}$ and $t^{\rm F}_{\rm p}$ is found. The best fit gives $\log w^{\rm F}=-0.32+1.01\log t^{\rm F}_{\rm p}$, i.e., $w^{\rm F}\sim t^{\rm F}_{\rm p}/2$. The $L^{\rm F}_{R,p}$ is anti-correlated with $t^{\rm F}_{\rm p}$ in the burst frame, i.e., $\log L^{\rm F}_{\rm R, iso,48}=(1.89\pm 0.52)-(1.15\pm0.15) \log [t^{\rm F}_{\rm p}/(1+z)]$ with a Spearman correlation coefficient of 0.85 and a chance probability $p<10^{-4}$. Therefore, a flare peaking at a later time tends to be dimmer and wider.

$E^{\rm F}_{\rm R, iso}$ are usually smaller than 1/100 of $E_{\gamma, \rm iso}$. The $L^{\rm F}_{\rm R, iso}$ is correlated with $L_{\rm \gamma, iso}$, i.e., $\log L^{\rm F}_{\rm R, iso}/10^{48}=(-3.97\pm 0.60)+(1.14\pm 0.27)\log L_{\rm \gamma, iso}/10^{50}$ with a Spearman correlation coefficient of $r=0.75$ and a chance probability $p\sim 10^{-3}$. The flares in GRBs 050401, 060926, and 090726 are out of the $3\sigma$ region of the fit. Without considering the flares in the three GRBs, it is found that the $t^{'\rm F}_{\rm p}$ is also tightly anti-correlated with $E_{\gamma, \rm iso}$ i.e., $\log t^{'\rm F}_{\rm p}=(5.38\pm 0.30)-(0.78\pm 0.09)\log E_{\rm \gamma, iso}/10^{50} $ (with $r=0.92$). Similarly, a tight anti-correlation between $L_{\rm R, p}$ and $t_{\rm p}$ in the burst frame is found without considering the flares in the three GRBs, e.g., $\log [t^{\rm F}_{\rm p}/(1+z)]=(7.57\pm 0.60)-(1.35\pm 0.17)\log E_{\gamma,\rm iso, 50}$ with a Spearman correlation coefficient of 0.91.

It is interesting to study whether the optical flares are associated with X-ray flares. Early optical flares are only observed in the lightcurves of GRBs 060210, 060926, 090618, and 090726 in our sample, indicating that the fraction of GRBs with detection of early optical flares is much lower than that of the X-ray flares. Among the 19 GRBs with detection of the optical flares 16 are detected with Swift. Simultaneous observations with XRT during the optical flares are available for GRBs 050401, 060206, 060210, 060607A, 060926, 070311, 071010A, 071031, 080506, 090618, and 100728B.  An X-ray flare that may be associated with the optical flare is only observed in GRBs 060926, 070311, and 071010A. The optical flares of the three GRBs are lagged behind the corresponding X-ray flares. Measuring the lags with the peak times of the flares, we get 196 seconds, $7.7\times 10^4$ seconds, and $2.45\times 10^4$ seconds for the flares in GRBs 060926, 070311, and 071010A, respectively. The lag is potentially  proportional to the peak time of the flares with the three flares.
\subsection{Prompt Optical and Reversed Shock Flare}
Well-sampled prompt optical flares are observed for GRBs 061121, 060526, 080129, 080319B, and 110215A. They usually trace the pulses of the prompt gamma-ray phase with a significant temporal lag. Reversed shock Flares are detected for GRBs 990123 and 060111B. Generally speaking, the light curve dominated by reverse the shock emission is fast rising and decaying, quite similar to the prompt optical flares.
\subsection{Late Flare}
%\begin{figure*}
%\includegraphics[angle=0,scale=0.350,width=0.3\textwidth,height=0.2\textheight]{050401.ps}
%\includegraphics[angle=0,scale=0.350,width=0.3\textwidth,height=0.2\textheight]{070311.ps}
%\includegraphics[angle=0,scale=0.350,width=0.3\textwidth,height=0.2\textheight]{080319B.ps}
%\caption{Optical flares observed in GRBs 050401, 0703110, and 80319B. The solid lines represent the best fit to the data. Simultaneous X-ray data
%observed with {\em Swift}/XRT (crosses with error bars) are also presented.}
%\label{Opt_LC_Flares}
%\end{figure*}
\section{Early Shallow Decay Segment}
\begin{figure*}
\includegraphics[angle=0,scale=0.350,width=0.32\textwidth,height=0.25\textheight]{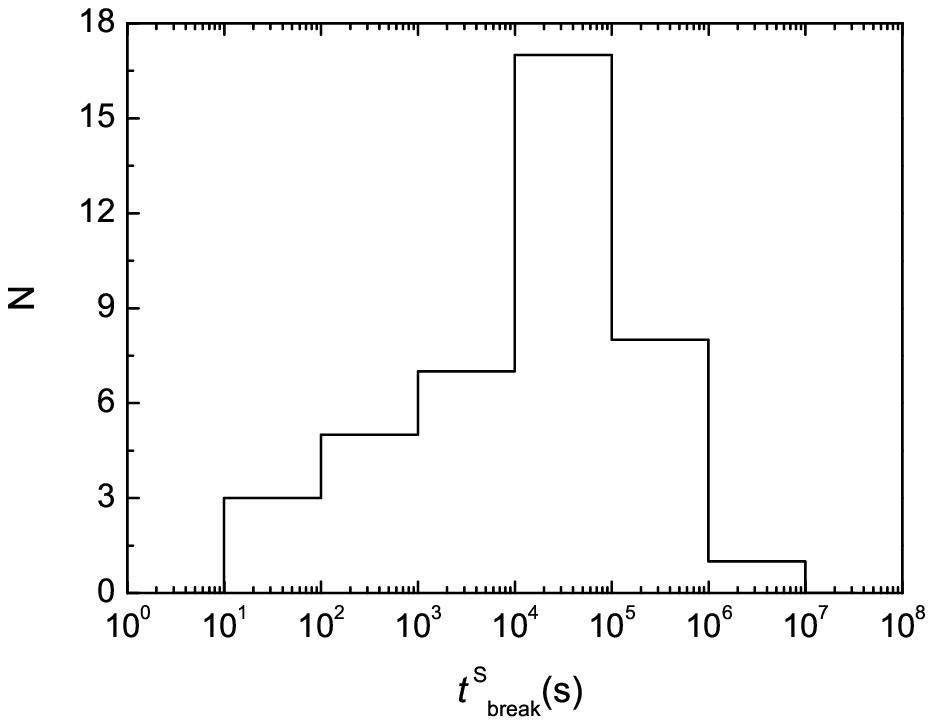}
\includegraphics[angle=0,scale=0.350,width=0.32\textwidth,height=0.25\textheight]{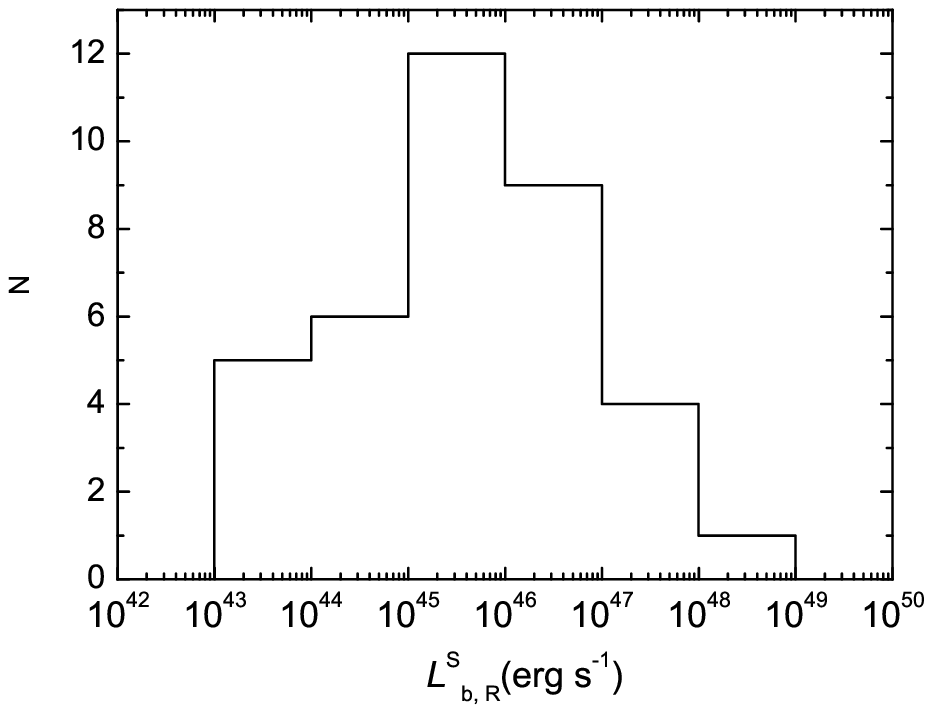}
\includegraphics[angle=0,scale=0.350,width=0.32\textwidth,height=0.25\textheight]{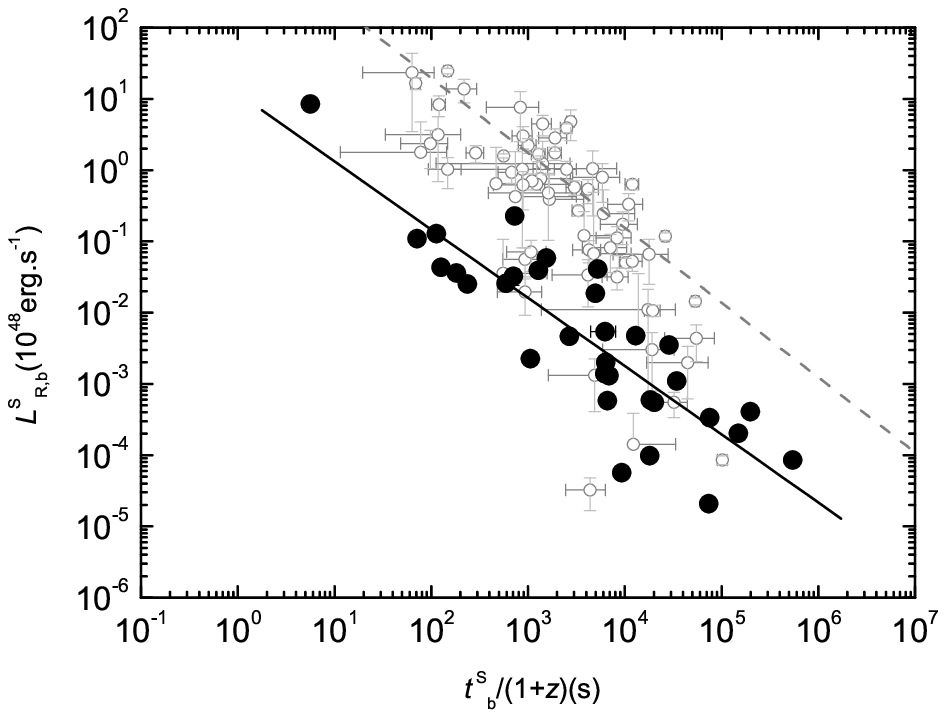}
\caption{Correlation and Distributions of $L^{S}_{\rm R, iso}$  and $t^{\rm S}_{b}/(1+z)$ of the shallow decay segments for the GRBs in our sample. The grey circles are for the the X-ray data from Dainotti et al. (2010). Lines are the best fit line.} \label{Shallow_corr}
\end{figure*}

We get a sample of 41 GRBs with a shallow decay segment from the 146 GRBs. Thirty-one out of the 41 shallow decay segments transit to a decay slope of $1\sim 2.5$, and 5 of them are followed by a sharp drop with a decay slope being steeper than 2.5. About half of the shallow decay segments look like a plateau, with $|\alpha^{\rm S}_{b,1}|\leq 0.3$. Figure \ref{Shallow_corr} shows the distributions of the break times ($t^{\rm S}_{\rm b}$) and the luminosity at the break ($L^{\rm S}_{b, \rm iso}$) as well as their correlation. The break time ranges from tens of seconds to several days post the GRB trigger, with a typical $t^{\rm S}_{\rm p}$ of $\sim 10^4$ seconds. The $L^{\rm S}_{\rm R,b}$ varies from $10^{43}$ to $10^{47}$ erg s$^{-1}$, and even $\sim 10^{49}$ erg s$^{-1}$ of the early break in some GRBs. The $L^{\rm S}_{\rm R,b}$ is anti-correlated with $t^{\rm S}_{\rm b}$, which is $\log L^{\rm S}_{\rm R,48}=(1.75\pm 0.22)-(0.78\pm 0.08)\log [t^{\rm S}_{\rm b}/(1+z)]$ with a Spearman correlation coefficient of $r=0.86$ and $\rho<10^{-4}$.

A shallow decay segment is commonly seen in the well-sampled of XRT lightcurves, except for a few GRBs that their XRT lightcurves decay as a single power-law\cite{Liang2009} (Liang et al. 2009). It was also reported that the X-ray luminosity at the break time is correlated with the break time \cite{Dainotti2010}(Dainotti et al. 2010). We over-plot the $L_{\rm b, iso}$ as a function of $t_{\rm b}$ in the burst frame in Figure \ref{Shallow_corr}. One can observe that optical data share the same relation to the X-ray data. Note that the X-ray luminosity is in the 0.3-10 KeV energy band and the optical data are in the R band, so that the X-ray luminosities are significantly higher than the optical ones. The observed photon indices of the X-ray spectra are $\sim 2$. Therefore, the energy spectra of the X-rays are flat and the derived $L_{\rm b, iso}-t_{\rm b}$ relation in the 1 KeV band is roughly consistent with that observed in the X-ray band.

\begin{figure*}
\includegraphics[angle=0,scale=0.350,width=0.32\textwidth,height=0.25\textheight]{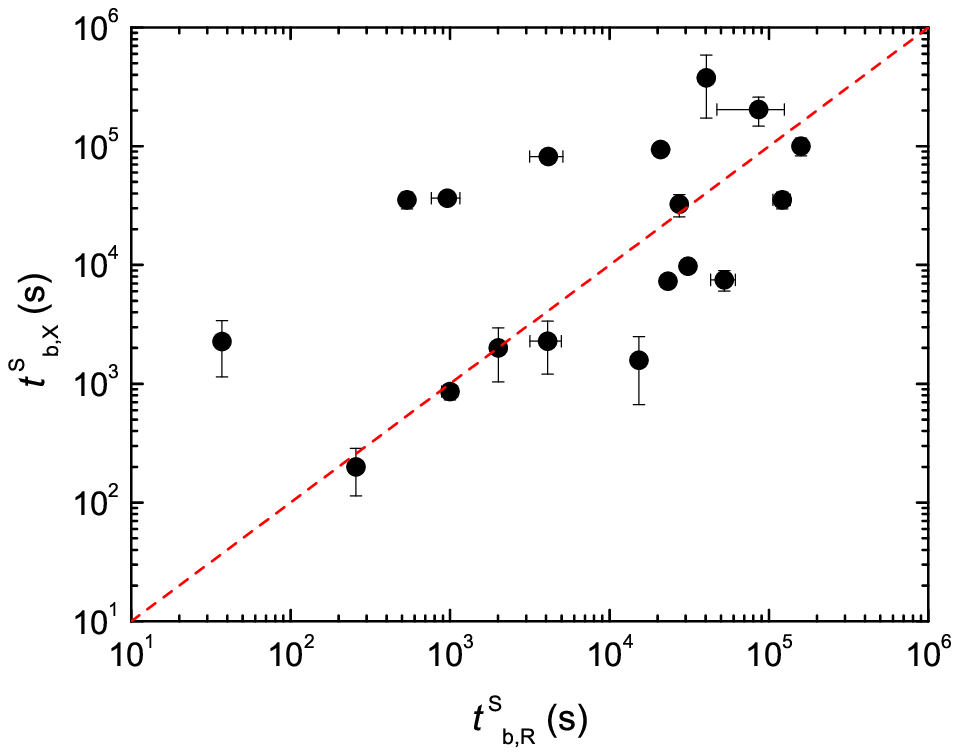}
\includegraphics[angle=0,scale=0.350,width=0.32\textwidth,height=0.25\textheight]{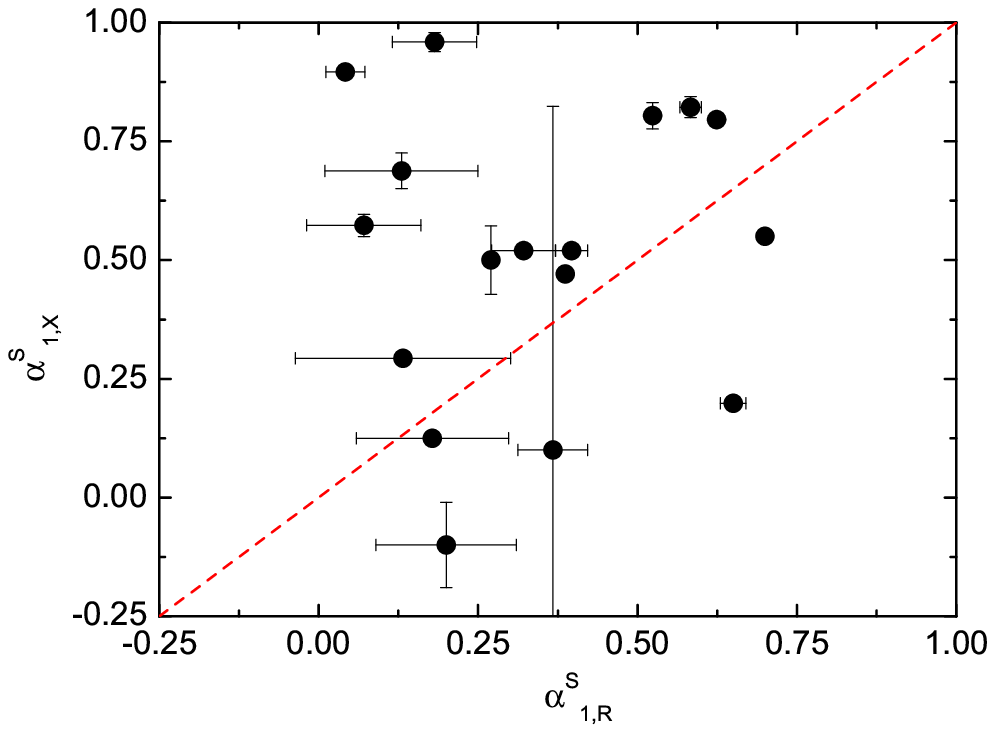}
\includegraphics[angle=0,scale=0.350,width=0.32\textwidth,height=0.25\textheight]{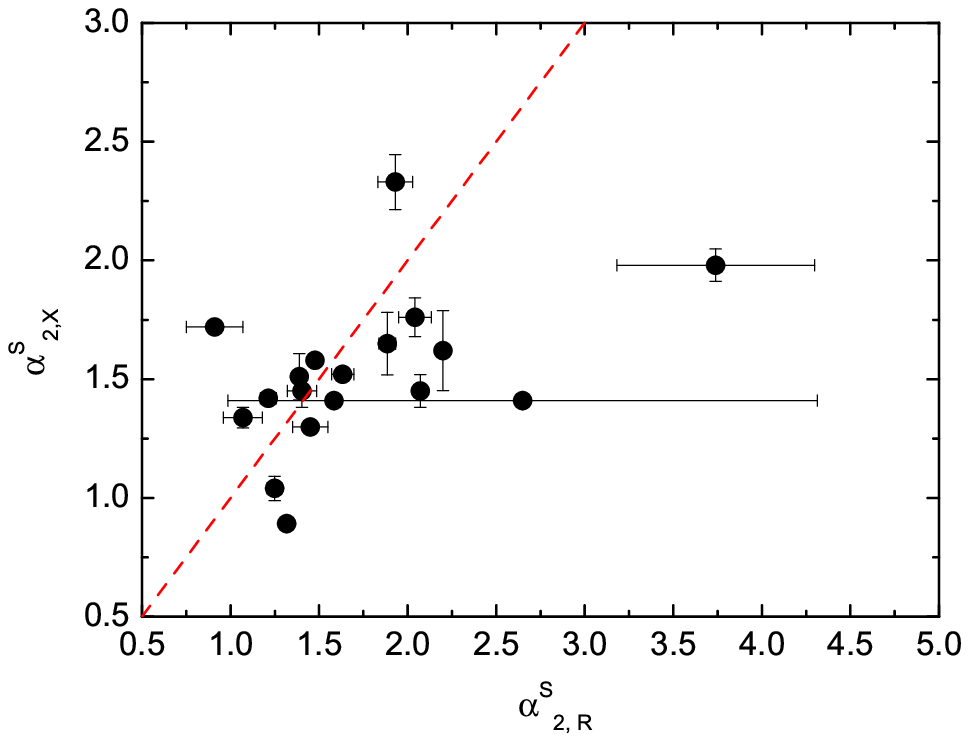}
\caption{Comparisons of the decay slopes and the break times in the optical and X-ray bands. The dashed lines are the equality lines. } \label{Shallow_Opt_Xray}
\end{figure*}

We examine the chromaticity of the shallow decay segments in the X-ray and optical bands.  The X-ray observations are available for 17 out of the 34 GRBs. We extract the underlaying afterglow components from the X-ray data and compare the $\alpha^{\rm S}_{\rm 1}$, $\alpha^{\rm S}_{\rm 2}$, and $t^{\rm S}_{\rm b}$ in the X-ray and optical bands in Figure \ref{Shallow_Opt_Xray}. It is found that the data points are scattered around the equality line, and a tentative correlation between the break times of the optical and X-ray lightcurves is observed, with a chance probability of the correlation of  $\sim 0.15$. These is no correlation between the decay slopes of the X-ray and optical lightcurves. The decay segment prior to the break times in the optical bands tend to be steeper than that of the X-ray band, but the slopes post the breaks are roughly consistent, except for those $\alpha^{\rm S}_{2}>2.5$ in the optical bands.
%\begin{figure*}
%\includegraphics[angle=0,scale=0.350,width=0.3\textwidth,height=0.2\textheight]{050922C.ps}
%\includegraphics[angle=0,scale=0.350,width=0.3\textwidth,height=0.2\textheight]{060605.ps}
%\includegraphics[angle=0,scale=0.350,width=0.3\textwidth,height=0.2\textheight]{080413B.ps}
%\caption{Examples of the optical lightcurves with detection of a shallow decay segment. The shallow decay segments in GRBs 060605 an d080413B seem to be a superimposed plateau. The symbols are the same as Figure 1.}
%\label{Shallow_LC}
%\end{figure*}

\section{Early Afterglow Onset Bump}
\begin{figure*}
\includegraphics[angle=0,scale=0.350,width=0.32\textwidth,height=0.25\textheight]{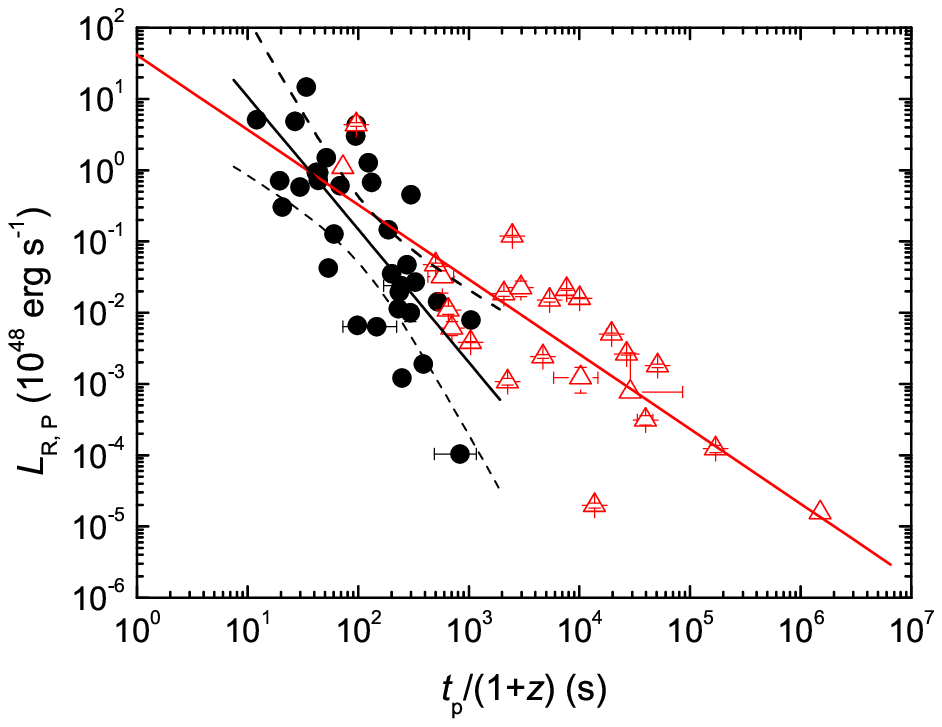}
\includegraphics[angle=0,scale=0.350,width=0.32\textwidth,height=0.25\textheight]{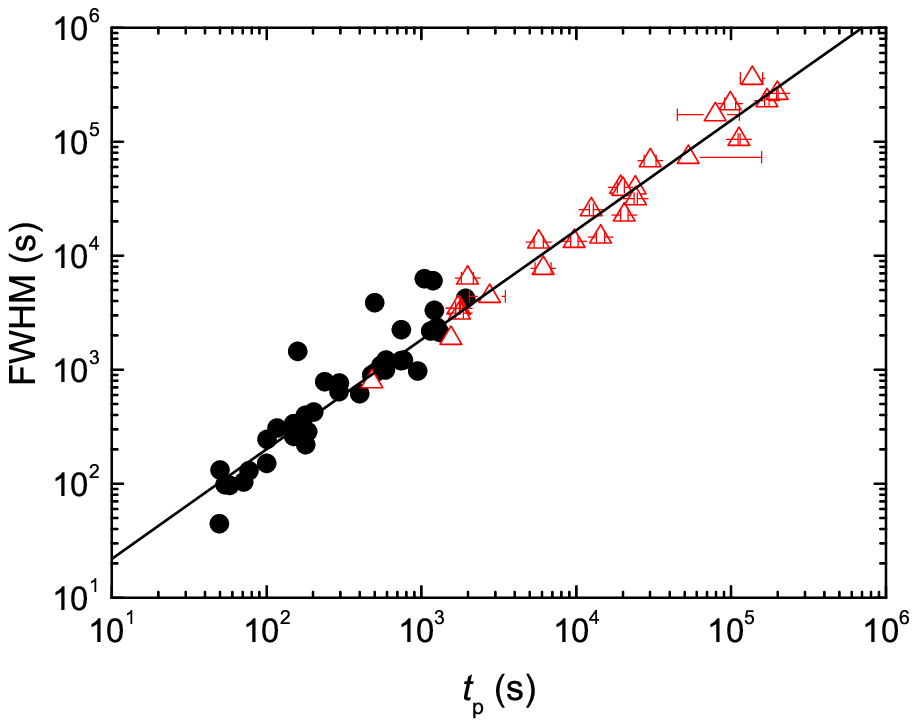}
\includegraphics[angle=0,scale=0.350,width=0.32\textwidth,height=0.25\textheight]{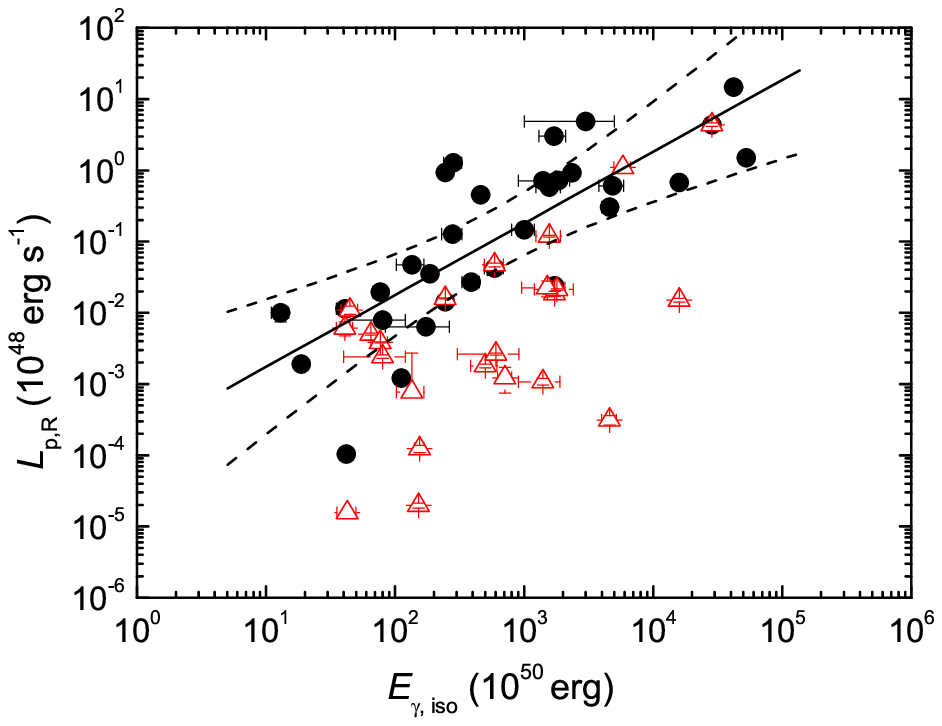}
\caption{Correlation between $w^{\rm A}$ and $t^{\rm A}_{\rm p}$ as well as relations of $L^{\rm A}_{\rm R, iso}$ to $t^{\rm A}_{\rm p}$ and $E_{\gamma, \rm iso}$. The black solid dots are for the afterglow onset bumps. The data for the late re-brightenings are also shown with opened triangles for comparison.}
\label{Onset_RB_Corr}
\end{figure*}

An early smooth bump is observed in the optical afterglow lightcurves of 42 GRBs in our sample. The peak luminosity and width as a function of $t^{\rm A}_{\rm p}$ are shown in Figure \ref{Onset_RB_Corr}. It is found that $L^{\rm A}_{\rm {R,p}}$ is anti-correlated with $t^{\rm A}_{\rm p}$ measured in the burst frame and $w$ is tight correlated with $t^{\rm A}_{\rm p}$, indicating that it is wider and dimmer if it peaks later. The isotropic prompt gamma-ray energy ($E_{\gamma, {\rm iso}}$) is tightly correlated with $L^{\rm A}_{\rm R,p}$ (Figure \ref{Onset_RB_Corr}). The best fit yields $L^{A}_{R, iso}\propto E_{\gamma, \rm iso}^{1.00\pm 0.14}$. Assuming that the bumps signal the deceleration of the GRB fireballs in a constant density medium, we calculate the initial Lorentz factor ($\Gamma_0$) of the GRBs with redshift measurements. The derived $\Gamma_0$ are typically a few hundreds. The $\Gamma_0-E_{\rm \gamma, iso}$ relation discovered by Liang et al. (2010)\cite{Liang2010} is confirmed with the current sample.
%
%\begin{figure*}
%\includegraphics[angle=0,scale=0.350,width=0.3\textwidth,height=0.25\textheight]{081203A.ps}
%%\includegraphics[angle=0,scale=0.350,width=0.3\textwidth,height=0.25\textheight]{100906.ps}
%\includegraphics[angle=0,scale=0.350,width=0.3\textwidth,height=0.25\textheight]{080810.ps}
%\caption{Examples of the optical lightcurves with detection of a clear afterglow onset phase. The symbols are the same as Figure 1.}
%\label{Onset_LC}
%\end{figure*}

\section{Late Re-Brightening}
A re-brightening hump is analogous to the afterglow onset hump but it is latter than the onset humps. It is observed in 30 GRBs in our sample. Both $\alpha_1$ and $\alpha_2$ of the onset and re-brightening bumps are well consistent. Most $\alpha_1^{,}$s are in the range of $-3\sim 0$, with a typical value of $\sim -1$. The $\alpha_1$ in some cases is smaller than $-3$. The $\alpha_2$ values are in the range of $0.3\sim 2$. It is found that the $t^{\rm R}_{\rm p}$ randomly ranges from several hundreds of seconds to several days. The $L^{\rm R}_{\rm p}$ is systematically lower than $L^{\rm A}_{\rm p}$. The typical $E^{R}_{\rm R, iso}$ is $10^{45}\sim 10^{47}$ erg. Correlation of the characteristics of the re-brightening bumps are also shown in Figure \ref{Onset_RB_Corr} in comparison with the afterglow onset bumps. It shares the same relation between the width and the peak time as the onset bump, but no clear correlation between $L_{R, p}$ and $E_{\gamma, iso}$, as is the case for the onset bumps. Although its peak luminosity also decays with time, the slope is much shallower than that of the onset peak. We get $L\propto t^{-1}_{\rm p}$, being consistent with off-axis observations of an expanding fireball in a wind-like circum medium\cite{Panaitescu2008}. Therefore, the late re-brightening may signal another jet component.

%\begin{figure*}
%\includegraphics[angle=0,scale=0.350,width=0.3\textwidth,height=0.25\textheight]{030329.ps}
%\includegraphics[angle=0,scale=0.350,width=0.3\textwidth,height=0.25\textheight]{060906.ps}
%\includegraphics[angle=0,scale=0.350,width=0.3\textwidth,height=0.25\textheight]{100901A.ps}
%\caption{Examples of the optical lightcurves with late re-brightening. The symbols are the same as Figure 1.}
%\label{RB_LC}
%\end{figure*}
\section{Summary}
We have analyzed a well-sampled lightcurves of 146 GRBs. A generic optical lightcurve and statistical results of various emission
components are presented. We summary our results below.

(1) Twenty-four late optical flares are obtained from 19 GRBs. The fraction of the detected optical flares is much smaller than that of X-ray flares. Associated X-ray flares are observed for 4 optical flares and the optical flares usually lag behind the corresponding X-ray flares. We find $L^{\rm F}_{\rm R, iso}\propto L_{{\gamma}, \rm iso}^{1.11\pm 0.27}$,   $w^{\rm F}\sim t^{\rm F}_{\rm p}/2$ and $L^{\rm F}_{\rm R, iso}\propto  [t^{\rm F}_{\rm p}/(1+z)]^{-1.15\pm0.15}$, indicating that the optical flares are correlated with the prompt gamma-ray phase and a flare peaking latter is wider and dimmer. These results suggest that the physical origin of the late optical flares could  be the same as the prompt gamma-ray phase and the temporal evolution from the GRB phase to late optical flares may signal the global evolution of the GRB central engine.

(2) A shallow decay segment is observed in 39 GRBs. The detection fraction of the optical shallow decay component is comparable to that in the X-ray band. The X-ray and optical breaks are usually chromatic, but a tentative correlation is found. Their break times ($t^{\rm S}_{\rm b}$) range from tens of seconds to several days post the GRB trigger, with a typical value of $\sim 10^4$ seconds. The break luminosity is anti-correlated with $t^{S}_{\rm b}$, $L^{\rm S}_{\rm R, b}\propto {[t^{S}_{\rm b}/(1+z)]}^{-0.78}$, similar to that derived from X-ray flares. The shallow decays / internal plateaus may be evidence of a long-lasting wind powered by the central engine. The injection behavior may be used to diagnose the nature of the central objects in the GRB central engine. Assuming that the behavior of the luminosity injected into the forward shocks evolves as $L=L_0t^{-q}$, we find that the long-lasting wind may be powered by a Poynting flux from a black hole via the Blandford-Znajek mechanism fed by fall-back mass or by the spin-down energy release of a magnetar after the main burst episode\cite{Rees1998,Dai1998,Sari2000,Zhang2001}. One critical issue to explain the shallow decay segment with the energy injection scenario is the chromatic breaks in the optical and X-ray bands. Mixing of different emission components may be the reason for the observed chromatic breaks of the shallow decay segment in different energy bands.

(3) An early smooth bump is observed in the optical afterglow lightcurves of 42 GRBs in our sample. It is found that $L_{\rm {R,p}}$ is anti-correlated with $t^{\rm A}_{\rm p}$ measured in the burst frame and $w^{\rm A}$ is tightly correlated with $t^{\rm A}_{\rm p}$, indicating that a dimer flare tends to peak later and be wider. The $E_{\gamma, {\rm iso}}$ is also tightly correlated with $L^{\rm A}_{\rm R,p}$. Assuming that the bumps signal the deceleration of the GRB fireballs in a constant density medium, we calculate the initial Lorentz factor ($\Gamma_0$ of the GRBs with redshift measurements. The derived $\Gamma_0$ are typically a few hundreds. The $\Gamma_0-E_{\gamma, \rm iso}$ correlation discovered by Liang et al. (2010)\cite{Liang2010} is confirmed with the current sample. The tight relation of the onset bumps to the prompt gamma-rays may open a window to investigate the radiation physics of GRB fireballs.

(4)A re-brightening hump is analogous to the afterglow onset hump but it follows an onset hump or a power-law decay segment. It is observed in 30 GRBs in our sample. It shares the same relation between the width and the peak time as for the onset bumps, but no clear correlation between $L_{R, p}$ and $E_{\gamma, iso}$ is found. Although its peak luminosity also decays with time, the slope is much shallower than that of the onset peak. We get $L\propto t^{-1}_{\rm p}$, being consistent with off-axis observations to an expanding external fireball in a wind-like circum medium. Therefore, the late re-brightening may signal another jet component.

%
%
%\begin{thebibliography}{1}
%\bibitem[Antonelli et
%al.(2006)]{\it 2006A&A456...509...A} Antonelli, L.~A., et al.\ 2006, \aap, 456, 509
%
%
%\bibitem[Castro-Tirado et al.(1999)]{\it 1999ApJ...511L..85C} Castro-Tirado,
%A.~J., et al.\ 1999, \apjl, 511, L85
%
%
%\bibitem[Cenko et al.(2011)]{\it 2011ApJ...732...29C} Cenko, S.~B., et al.\
%2011, \apj, 732, 29
%
%
%
%
%
%\end{thebibliography}

\end{document}